\documentclass{webofc}
\usepackage[varg]{txfonts,epsfig}   
\woctitle{QCD@Work 2014}
\begin{document}
\title{Hadronization systematics and top mass reconstruction}
%
%

\author{Gennaro Corcella\inst{1}\fnsep\thanks{\email{gennaro.corcella@lnf.infn.it}} 
}

\institute{INFN, Laboratori Nazionali di Frascati, Via E.~Fermi 40, I-00044,
Frascati (RM), Italy}

\abstract{%
I discuss a few issues related to 
the systematic error on the top mass measurement at hadron colliders,
due to hadronization effects.
Special care is taken about the impact of bottom-quark fragmentation in top decays,
especially on the reconstruction relying on final states with leptons and $J/\psi$
in the dilepton channel.
I also debate the relation between the measured mass and its theoretical definition,
and report on work in progress, based
on the Monte Carlo simulation of fictitious top-flavoured hadrons,
which may shed light on this issue and on the hadronization systematics.}
\maketitle
\section{Introduction}
\label{intro}
Heavy-flavour, and in particular
top-quark phenomenology is nowadays one of the most lively
fields of investigation, in both theoretical and experimental particle physics.
At hadron colliders, top-quark pairs are
mainly produced through strong interactions and decay 
via $t\to bW$ before hadronizing; the final states are then classified as
dileptons, leptons+jets or all-jets, according to the decay mode of the $W$ boson.
The top-quark mass is then measured by reconstructing its decay 
products. 

In fact, the top-quark mass $m_t$ is a fundamental parameter
of the Standard Model, as it has been employed, together with the $W$ mass,
even before the discovery of
the Higgs boson \cite{higgs},
in the electroweak precision tests constraining the Higgs mass \cite{baak}.
Furthermore, using its current world average, i.e. 
$m_t=173.34\pm 0.27$ (stat) $\pm$ 0.71 (syst)~GeV \cite{wav}, 
and the Higgs mass results,
one finds that the Standard Model vacuum lies at the border between
stability and metastability regions \cite{gino}.
For the above reasons, it is clear that it is
necessary to have under control all possible sources of uncertainties on
the top mass determination, such as the theory systematics, and 
that it would be desirable to measure $m_t$
with the highest possible precision. 

In this paper I investigate 
the uncertainties on the measurement of the top mass, paying special attention to the
hadronization corrections, such as the modelling of bottom-quark 
fragmentation in top decays. In fact, $b$-quark fragmentation enters in the
theoretical uncertainty on the top mass reconstruction, as it contributes to 
the Monte Carlo systematics.
The world average determination, 
based on standard reconstruction methods, such as the template, ideogram and
matrix-element techniques.
quotes an overall Monte Carlo uncertainty of 380 MeV,
whereas the systematic error due to $b$-tagging and $b$-jet energy scale,
also depending on bottom fragmentation in top decays , are about 
250 and 110 MeV, respectively.
An even larger impact of the treatment of the hadronization mechanism
is found in the method relying on final states where both 
$W$'s decay leptonically ($W\to\ell\nu$) and the $B$ hadron into a $J/\psi$
\cite{avto}.
Although the decay $B\to J/\psi X$ is a rare one, the events where the $J/\psi$ decays via
$J/\psi \to\mu^+\mu^-$ give a fully leptonic final state, in such a way that,
after setting suitable cuts on lepton rapidities and transverse momenta, 
a measurement of the peak value of the three-lepton invariant mass
$m_{3\ell}$ or of the $J/\psi\ell$ one, i.e. $m_{J/\psi\ell}$, allows a 
reliable fit
of $m_t$. The total systematic error
is $\Delta m_t\simeq 1.47$~GeV \cite{chierici}, with 1.37 GeV
due to the theoretical uncertainties;
the contribution of $b$-quark fragmentation to the systematics,
estimated by running the PYTHIA event generator \cite{pythia}, 
is about 0.7 GeV.
Refs.~\cite{volker} and \cite{corme} further investigated 
this issue and compared PYTHIA with HERWIG \cite{herwig}, 
after fitting their hadronization models to LEP and SLD data: it was then
stated that the actual uncertainty can be even larger than what
estimated in \cite{avto,chierici} using only one event generator.

In this paper I address some topics which are relevant for
the sake of a reliable estimate of the theory uncertainty on the
top mass extraction.
In Section 2, I reconsider the findings of \cite{volker,corme} and
compare the Monte Carlo predictions with the NLO calculation 
\cite{nlo} as well.
In Section 3 I briefly discuss the relation between
reconstructed mass and theoretical definitions, such as the pole mass, 
and present some ideas to address this issue, based on the Monte Carlo
simulation of fictitious top-hadron states.
Section 4 contains some concluding remarks.

\section{Bottom-fragmentation in top decays and impact on the theory systematics}
Bottom-quark fragmentation in top decays has been thoroughly investigated through the
years, according to fixed-order and resummed calculations,
as well as Monte Carlo event generators.
The total top-quark width has been calculated up to next-to-next-to-leading
order (NNLO) accuracy in \cite{cz}, whereas the $b$-quark energy
spectrum has been obtained by means of soft/collinear 
resummed calculations, based on the perturbative fragmentation
approach \cite{mele}, to next-to-leading-logarithmic (NLL) accuracy \cite{cormit,ccm}
and matched to the exact NLO result.
Hadronization effects are included by convoluting the resummed
$b$-quark distribution with a non-perturbative fragmentation function,
such as the Kartvelishvili or Peterson models.
An alternative, and perhaps theoretically better approach, 
first advocated in \cite{cn} and later used
even in \cite{ccm}, consists of
working in Mellin moment space, with a $N$-space non-perturbative
fragmentation function, without making any assumption
on its functional form in $x$-space.

Ref.~\cite{nlo} computed the invariant mass distribution
of a charged lepton, coming from $W$ decay, and a $B$ meson at
NLO, using the fits in \cite{cormit} to model
non-perturbative corrections.
NNLO distributions in top decays, with massless $b$ quarks and without accounting
for hadronization effects, were presented in \cite{caola,gao}.
While all the above computations have been undertaken in the
narrow-width approximation with on-shell top quarks, Ref.~\cite{denner}
took into account interference effects between top production
and decay at NLO and accounted for the resummation of some enhanced contributions
by using a dynamical renormalization scale, depending on the
top-quark transverse momentum. 
Of course, it would be very
interesting to match the parton-level calculations \cite{caola,gao,denner}
with suitable hadronization models.

As far as Monte Carlo generators, such as HERWIG or PYTHIA, 
are concerned, they include radiative effects
by means of parton showers, equivalent to a resummation of leading
soft/collinear logarithms (LLs), with the inclusion of some NLLs
\cite{cmw}. The
transitions of partons to hadrons is then modelled via the string \cite{string}
(PYTHIA) and cluster (HERWIG) \cite{cluster} models.
All non-perturbative models depend on a few free parameters which are 
be tuned to experimental data; although using the LHC data would be of course 
desirable, at the moment the best fits employ
$e^+e^-\to b\bar b$ data from LEP and SLD experiments.
The best-fit parametrizations can then be used to predict $B$-hadron production in top
decays at the LHC, paying attention to the consistency of the procedure, i.e.
tunings relying on resummed computations or parton showers
for $e^+e^-$ annihilation must be used only if $b$-quark ($B$-hadron) 
production in top decay is treated in the same manner. 
When using a formalism entirely developed in moment space, one can fit
directly the moments of the non-perturbative fragmentation function
and use such moments for other processes,
like top decays.
LEP and SLD collaborations 
did tune hadronization models to the $B$-hadron energy
spectrum at the $Z$ pole, but every 
fit yielded different numbers, so that one could not make any
final statement on a parametrization which could eventually be 
implemented into a Monte Carlo code. 

Ref.~\cite{volker} presented an improved tuning of bottom fragmentation,
since OPAL \cite{opal}, ALEPH \cite{aleph}
and SLD \cite{sld} data were taken into account altogether,
as if they came from the same experimental sample.
The finding is that the default versions of PYTHIA 6 and HERWIG 6 were 
unable to describe $B$-hadron data ($\chi^2/{\rm dof}\sim {\cal O}(10)$ 
for both codes), but one had to retune the cluster and string models.
After the tuning, PYTHIA managed to give a good fit of the data
($\chi^2/{\rm dof}\simeq 1$), while HERWIG, although much improved
with respect to the default parametrization,
 still exhibited meaningful discrepancies ($\chi^2/{\rm dof}\simeq 3.5$) .
The drawback of this analysis lies in the fact that only one observable,
namely the $B$-hadron energy fraction in the $Z$ rest frame $x_B$,
was tuned and therefore the agreement with other measurements, e.g.
involving light hadrons, might have been spoiled. As for PYTHIA, 
the universality of the $a$, $b$ and $r$ parameters of the Lund model
\cite{string} is clearly not
respected in the parametrization proposed in \cite{volker}. 
The tuning of \cite{volker} can thus be improved
by choosing values of $a$, $b$ and $r$ which respect the universality of
the Lund model and are capable of reproducing both heavy- and light-flavoured
hadron data \cite{talk}.
Also, tuning CLSMR(1), the gaussian
smearing for light-hadron production with respect to the
direction of the parent quark, was found not to be essential
when dealing with top and bottom quarks \cite{talk}.
Though with these limitations, the fits in \cite{volker} 
were nonetheless used in the jet-energy measurement by
the ATLAS Collaboration \cite{atlas}. 

The tuning in \cite{volker} was also employed in \cite{corme},
where, extending previous analyses undertaken in
\cite{corc,cms}, 
HERWIG and PYTHIA were used to study the
$B$-lepton invariant-mass distribution for $t\bar t$ events at the LHC, 
in the dilepton channel, and possibly
extract $m_t$ from the measurement of the Mellin moments of $m_{B\ell}$.
The $m_{B\ell}$ spectrum is in fact quite interesting, since
it is Lorentz-invariant, thus roughly independent of
initial-state radiation and $t\bar t$ production mechanism,
and can be used to obtain, after a convolution with the $B\to J/\psi X$
one, the $m_{J/\psi\ell}$ invariant mass distribution and then obtain
the top mass as explained in \cite{avto,chierici}.
Eq.~(\ref{mtfit}) presents the linear fits which parametrize
the average value $\langle m_{B\ell}\rangle$ in terms of $m_t$, 
according to HERWIG (HW), PYTHIA (PY), tuned as in \cite{volker},
and the NLO calculation of \cite{nlo}, 
carried out within the perturbative fragmentation formalism, taking
the non-perturbative information from \cite{cormit}. 
\begin{eqnarray}
\langle m_{B\ell}\rangle_{\mathrm{HW}} &\simeq &  
-25.31~\mathrm{GeV} +0.61\  m_t \  ;\ \delta = 0.043~\mathrm{GeV};\nonumber\\
\langle m_{B\ell}\rangle_{\mathrm{PY}} &\simeq & 
-24.11~\mathrm{GeV} +0.59\  m_t\ ;\ \delta = 0.022~\mathrm{GeV};\nonumber\\
\langle m_{B\ell}\rangle_{\mathrm{NLO}} &\simeq & 
-26.70~\mathrm{GeV} +0.60\  m_t \ ;\ \delta = 0.004~\mathrm{GeV}.
\label{mtfit}
\end{eqnarray}
\begin{figure}
\begin{center}
\resizebox{0.5\textwidth}{!}{%
  \includegraphics{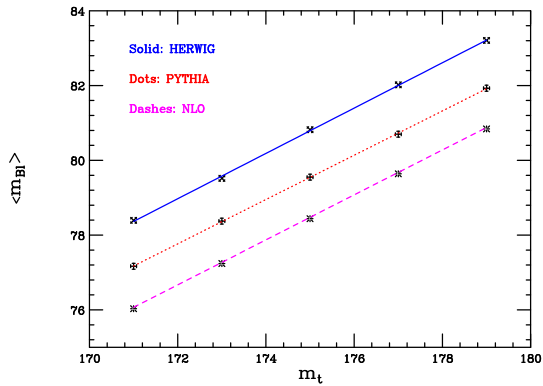}}
\caption{Average value of the invariant mass $m_{B\ell}$ as a function
of the top mass, according to HERWIG, PYTHIA and the NLO calculation,
tuned to $e^+e^-$ data from LEP and SLD.}
\label{fig:1} 
\end{center}      
\end{figure}
In (\ref{mtfit}), $\delta$ is the mean square deviation in the fit and
the straight lines are plotted in Fig.~\ref{fig:1}, in the range
171 GeV~$<m_t<$~179 GeV.
As Fig.~\ref{fig:1} exhibits, the three predictions are quite different and
can lead to a discrepancy $\Delta m_t$  between 1 and 2 GeV 
if $\langle m_{B\ell}\rangle$ were to be used to reconstruct $m_t$.
This difference, beyond the intrinsic differences in the PYTHIA,
HERWIG and NLO approaches, is mostly driven by the fact that the fits
of HERWIG to the $e^+e^-$ data were not very satistactory.
In perspective, using the object-oriented PYTHIA 8 \cite{py8}
and HERWIG++ \cite{herpp} codes can be a valuable
strategy for a reliable description of bottom-quark fragmentation
in the framework of Monte Carlo generators,
since both C++ programs manage to reproduce quite 
successfully the $e^+e^-$ data, with parametrizations
which work well even for light hadrons \cite{konrad}.

\section{Reconstructed mass and theoretical definitions}

The reconstructed top mass, based on the template, ideogram and matrix-element 
techniques, or even the lately proposed endpoint \cite{end} and $J/\psi$
\cite{chierici} methods, is obtained by comparing data with predictions yielded by
Monte Carlo simulations 
(see, e.g., Ref.~\cite{snowmt} for an update).
Since such codes are not exact NLO calculations,
the reconstructed mass cannot be precisely 
identified with any theoretical definition, such as the pole or 
$\overline{\rm MS}$ masses. 
The improved POWHEG \cite{powheg} and MC@NLO \cite{mcnlo}  
programs, used, e.g., in the analysis \cite{frix} to extract
the top mass from leptonic observables in the dilepton channel, 
do yield
the total NLO cross section, but still rely on HERWIG
or PYTHIA for showers and hadronization. 
In particular, generators like HERWIG and 
PYTHIA factorize top production and decay, thus neglecting interference
and width effects, which should instead be accounted in a full NLO
computation.
Therefore, one usually refers to the measured 
$m_t$ as a so-called `Monte Carlo mass'.

However, the very fact that $m_t$ is determined from
the kinematics of top-decay products, assuming on-shell top quarks, 
should lead to the conclusion that it must be close to the pole mass. 
In fact, analyses carried out in the framework of Soft Collinear Effective Theory
for $e^+e^-\to t\bar t$ annihilation, wherein
the Monte Carlo top mass can be associated with the jet mass 
evaluated at a scale of the order ot the shower cutoff,
i.e. about 1 GeV, have argued that the reconstructed mass differs from the 
pole mass by an amount $\sim {\cal O}(\alpha_S\Gamma_t)$
\cite{hoang}, $\Gamma_t$ being the top width.
Furthermore, the comparison of the measured $t\bar t$
cross section \cite{cross}
with the complete NNLO+NNLL calculation in \cite{czak},
making use of the pole mass, 
has proved that the pole mass is indeed compatible with the one determined
by means of the standard final-state reconstruction.
Nevertheless, the error on $m_t$, when extracted from the cross section
measurement, is still too large to make this method, albeit very interesting and
theoretically consistent, really competitive with respect to the
template or matrix-element techniques \cite{cross}.

\begin{figure}\hspace{1.cm}
{\epsfig{file=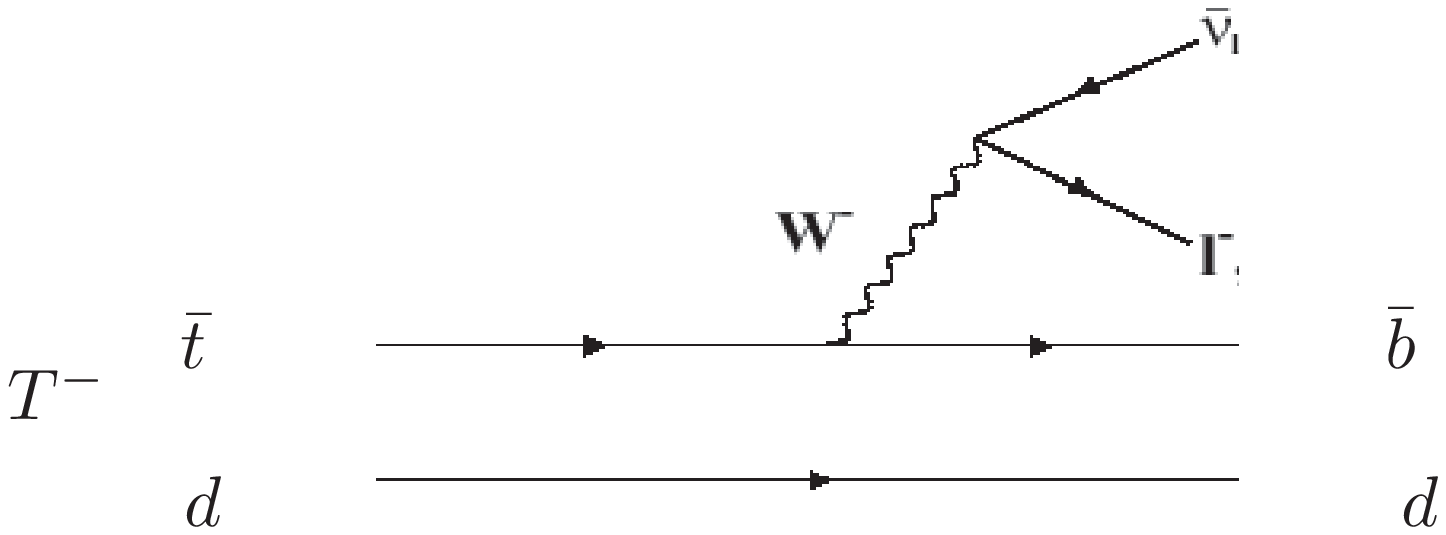,
angle=90,angle=90,angle=90,height=1.7in,width=6.in}}
\caption{Diagram with a fictitious top-flavoured hadron $T^-$ decaying according to the
spectator model.}
\label{fig:2}       
\end{figure}

In order to tackle the issue of relating Monte Carlo and pole masses,
a new investigation, based on the simulation of fictitious 
top-flavoured hadrons, was recently proposed \cite{talk}. 
In fact, if a quark is confined in a hadron of a given mass,
it is well known how it is possible to determine its mass,
in both on-shell and $\overline{\rm MS}$ renormalization
schemes, through, e.g., lattice methods, Non Relativistic QCD or
Heavy Quark Effective Theories. 
In fact, one can modify the shower and hadronization 
models in HERWIG or PYTHIA in such a way that the top quark
hadronizes, e.g., in  $T^0=t\bar u$ or $T^-=\bar t d$ mesons,
which then decay according to the spectator model.
In such processes, the light quark is the spectator one and the active $t$-quark
undergoes a $t\to bW$ transition, as displayed in Fig.~\ref{fig:2}
for a $T^-$ meson;
the $b$ and light quark possibly radiate further
gluons and finally hadronize.
One can then study final-state distributions for this sample of events,
such as the $B\ell$ invariant mass discussed above,
and reconstruct the top mass from its Mellin moments. 
If one compares such distributions with those obtained in
standard events, i.e. top quarks decaying without
hadronizing, one should be able to relate the Monte Carlo mass
with the $T$-meson mass and eventually with pole or
$\overline{\rm MS}$ masses. 
This investigation can therefore be a useful
benchmark to address the nature
of the Monte Carlo mass and the hadronization systematics. 
Moreover, the simulation of hadronized top quarks can also give some 
insight on the impact of colour-reconnection effects. In standard samples,
the top quark gets its colour from an initial-state gluon or light quark
and gives it to the bottom quark.
When hadronizing, it must instead create a colour singlet with 
the light (spectator) quark, which may yield a different colour flow in the
whole event.

Preliminary results of this analysis, already presented in \cite{talk} 
by using the HERWIG event generator, are quite cumbersome.
Above all, it was found that, in the hadronized samples, 
the spectator quark hardly radiates and there is even a non-negligible
fraction of events, proportional to the shower
Sudakov form factor \cite{cmw}, where the $b$ quark does not emit any gluon
either.
When this is the case, typically the $b$ quark forms a cluster of
small invariant mass, thus capable of decaying just into a $B$ meson plus
a soft hadron, e.g. a soft pion.
On the contrary, in standard top decays, the bottom quarks can radiate
several soft/collinear gluons and give rise to quite energetic
clusters.
Considering the $BW$ invariant
mass, as a toy-observable sensitive to the top-quark mass, 
its spectrum is in general quite different according to whether the top
quark hadronizes or not.
Selecting only those events where in
both simulations the $b$ quark emits gluons will make the 
$m_{BW}$ spectra look alike. 
Table~\ref{tab:1} presents the 
the Mellin moments of $m_{BW}$ for the events with gluon radiation
from the $b$ quarks as a function of the Monte Carlo top mass
in the range $171~{\rm GeV}<m_t<179~{\rm GeV}$.
One can learn that the Mellin moments of $m_{B\ell}$ when the top quark is forced to
hadronize before decaying are slightly larger than those in the standard samples,
by an amount which runs from 1\% ($\langle m_{B\ell}\rangle$) to 4\%  
($\langle m_{B\ell}^4\rangle$).
\begin{table}
\caption{First four Mellin moments of the $m_{BW}$ invariant mass
distribution yielded by HERWIG with hadronized top quarks. In brackets,
the moments for standard $t\bar t$ events are quoted.}
\label{tab:1} 
\begin{tabular}{lllll}
\hline\noalign{\smallskip}
$m_t$ (GeV) & $\langle m_{BW}\rangle$ (GeV)& 
$\langle m_{BW}^2\rangle$ (GeV$^2$)&
$\langle m_{BW}^3\rangle$ (GeV$^3$)& $\langle m_{BW}^4\rangle$ 
(GeV$^4$)\\
\hline
171 & 148.76~(148.08) & $2.24~(2.21)\times 10^4$ & $3.41~(3.35)\times 10^6$ & 
$5.24~(5.11)\times 10^8$ \\
\hline
173 & 150.44~(149.56) & $2.29~(2.26)\times 10^4$ & $3.53~(3.46)\times 10^6$ & 
$5.48~(5.32)\times 10^8$ \\ 
\hline
175 & 152.18~(151.00) & $2.35~(2.30)\times 10^4$ & $3.66~(3.56)\times 10^6$ & 
$5.74~(5.54)\times 10^8$ \\
\hline
177 & 153.80~(152.60) & $2.40~(2.36)\times 10^4$ & $3.77~(3.67)\times 10^6$ & 
$5.99~(5.78)\times 10^8$ \\ 
\hline
179 & 155.61~(153.97) & $2.45~(2.40)\times 10^4$ & $3.91~(3.78)\times 10^6$ & 
$6.28~(6.00) \times 10^8$ \\
\noalign{\smallskip}\hline
\end{tabular}
\end{table}
Such results, though quite stimulating, are not conclusive and
need further investigation: for example,  
rather than a $B$ meson,
one should consider the jet containing the $B$ meson and apply
the same jet-clustering algorithms as the experimental
analyses.
Also, it will be very interesting investigating how much the numbers in
Table~\ref{tab:1} depend on the shower cutoff and whether it may be possible
modifying the decay model, e.g. by means of a Lorentz boost, in such a way that
even the spectator quark is allowed to radiate. 
This is in progress.

\subsection{Conclusions}
I investigated some contributions to the
hadronization systematics on the top mass measurement,
namely the impact of the bottom-quark fragmentation 
and non-perturbative effects in the definition of the
reconstructed top mass. HERWIG and PYTHIA, the two most popular
parton shower and hadronization models, were retuned to $e^+e^-\to b\bar b$ data
from LEP and SLD, but nevertheless, even after the fit, they exhibit 
some relevant discrepancies.
Differences are also observed when comparing with
NLO calculations for quantities relying on top decays, such as the $B$-lepton
invariant mass in the dilepton channel. 
This finding is clearly a signal that 
bottom fragmentation is an issue which has to be further tackled 
to make a statement on its contribution to the uncertainties on
the top quark mass. As for the Monte Carlo codes, using
the object-oriented HERWIG++ and PYTHIA 8, improving both
shower and hadronization models,  is mandatory in
order to understand better the difference with respect to the NLO
computation. Also, a direct measurements of $b$-fragmentation in
top decays at the LHC will allow 
a direct fit of the hadronization models and therefore a test of factorization as well,
after comparing
with the results in $e^+e^-$ annihilation.

I also discussed the relation between the measured so-called
Monte Carlo mass and theoretical definitions like the
pole mass. 
Since the top mass is reconstructed from the kinematic properties of
final-state leptons and jets, using Monte Carlo programs which
neglect interference effects between production and decay phases
should lead to the measurement of a quantity close to the pole mass,
as this is defined in the on-shell renormalization scale. 
In fact, calculations carried out in the framework of
Soft Collinear Effective Theory have displayed that the Monte Carlo
mass, determined as the jet mass at a scale given by the
shower cutoff, differs from the
pole mass by an amount of ${\cal O}(\alpha_S\Gamma_t)$.
Investigation of Monte Carlo samples with hadronized top quarks,
wherein the Monte Carlo mass turns to be a meson mass,
may be a useful benchmark and ultimately shed some light on this issue.
This is in progress as well.

\end{document}